\documentstyle{europhys}

\def\And{{\rm and\ }}

\def\stars{\bigskip\centerline{***}\medskip}

\newif\ifboo \boofalse

\def\Review#1{\boofalse{\it #1},}
\def\Name#1{{\sc #1},}
\def\Vol#1{\ifboo Vol. {\bf #1}\else{\bf #1}\fi}
\def\Year#1{\ifboo #1\else(#1)\fi}
\def\Book#1{\bootrue{\it #1},}
\def\Page#1{\ifboo {\rm p. #1}\else{\rm #1}\fi}

\begin{document}
\def\beqn{\begin{eqnarray}}
\def\eeqn{\end{eqnarray}}
\def\beqns{\begin{eqnarray*}}
\def\eeqns{\end{eqnarray*}}
\def\beq{\begin{equation}}
\def\eeq{\end{equation}}
\def\bea{\begin{array}}
\def\ea{\end{array}}
\def\one{1\hskip-.37em 1}
\def\<{\langle}
\def\>{\rangle}
\def\ham{\hat{H}}
\euro{}{}{}{}
\Date{July 1999}
\shorttitle{S. Roche Quantum interferences in quasicrystals}
\title{Quantum interferences in quasicrystals}
\author{Stephan Roche}
\institute{Department of Applied Physics, University of Tokyo,
 7-3-1 Hongo, Bunkyo-ku, Tokyo 113-8654, Japan}
\rec{}{}
\pacs{
\Pacs{71}{23Ft}{Quasicrystals}
\Pacs{72}{15$-$v}{Electronic conduction in metals and alloys} 
\Pacs{72}{10$-$d}{Theory of electronic transport; scattering mechanisms}
      }
\maketitle

\begin{abstract} 
Contributions of quantum interference effects occuring in 
quasicrystals are reported. First conversely to 
 metallic systems, quasiperiodic ones are shown to enclose
 original alterations of their conductive properties 
 while downgrading long range order. Besides, origin
 of localization mechanism are outlined within the context 
 of the original metal-insulator transition (MIT) found in these materials.
\end{abstract}

\section{Introduction}

Despite sustained effort and concern, today's understanding of exotic 
electronic properties of quasicrystals \cite{AlMn} remains unsatisfactory 
although quasicrytalline materials have already been implemented to 
miscellanous concrete applications\cite{QC-app,QC-app2}. In particular, the role of quasiperiodic order on 
electronic localization and transport is believed to genuinely entail the most 
unexpected experimental features whereas so far, no coherent theoretical 
framework has been successfully ascertained\cite{Berger,RocheJMP,Takeo,RocheFS}. As 
a matter of fact, one of the unprecedented tendency of quasicrystals
 is the enhancement of their conductive ability upon increasing 
contribution of static (structural disorder) or dynamical excitations 
(phonons). This has been strongly supported by many experimental 
evidences\cite{Berger} and is often refered as an original 
property in the litterature. Notwithstanding theoretically, 
given heuristical arguments\cite{RocheJMP,Sire} and numerical 
investigations (e.g. the Landauer conductance for quasiperiodic Penrose lattices 
\cite{Fujiwara-Land} or Kubo formula for 3D-quasiperiodic models \cite{RochePRL1}
) yield to uncomplete understanding of the observed properties which 
range from anomalously metallic behaviors to insulating ones 
\cite{Julien}. It is generally assumed that a specific ``geometrical localization process'' takes place in quasicrystals 
(sustained by critical states\cite{Kho,Macia}) and that local disruptions of 
corresponding mesoscopic order reduce quantum interferences, resulting 
in an increase of conductivity. To tackle this issue on 1D 
quasiperiodic potential, tight-binding (TBM) as well as
 continuous Kronig-Penney models have been considered \cite{RocheFas}, 
 and phason-type disorder were shown to
disclose manifestations of quantum interferences in quasiperiodic order. 
Here we will show that alteration of critical states may under certain 
circumstances entail alteration of their propagating ability.
Besides, the role of quantum interferences on both sides of the
 quasicrystalline MIT for higher dimensional materials 
 is outlined.

\section{Anomalous electronic conductance in 1D-quasicrystals}

For 1D-quasiperiodic systems, we define phasons \cite{RocheFas}
that keep the essential characteristic of real systems, in the 
sense that they are a generic form of disorder which has no equivalent 
in usual metallic and periodic systems. Introducing random disorder into the 1D 
quasicrystal yield to Anderson localization in the infinite limit. For 
finite systems, localization lengthes may be much larger than the 
characteristic size so that conductance fluctuations as a function of 
energy of tunneling electrons (from the leads to the system) keep its 
self-similar nature and still follow power law 
behavior\cite{Dasarma}. It is thus difficult to relate localization 
and transport in such situations. On the other hand, tight-binding models (TBM) of perfect quasiperiodic 
chains have been intensively worked out both analytically and numerically only for some
given energies, but the results are supposed to have provided typical features of 
localization in quasiperiodic structures, such as power-law decrease 
of wavefunctions or power-law bounded resistances \cite{Kho}. Let us 
investigate the role of phasons in TBM. Hamiltonian is
${\cal H}=\sum_{n}t_{n}(|n\>\<n+1|+|n+1\>\<n|)$ 
($\gamma =t_{A}/t_{B}$ will stand for intensity of quasiperiodic 
potential, following a Fibonacci sequence ABAABABAABAAB\ldots, whereas site energies are kept constant)  and the 
Schr\"{o}dinger equation on a localized basis gives

\beq
\left(
\bea{c}
\psi_{n+1} \\
\psi_{n} \\
\ea
\right)=M_{n}
\left(
\bea{c}
\psi_{n} \\
 \psi_{n-1} \\
\ea
\right)=M_{n}M_{n-1}....M_{1}
\left(
\bea{c}
\psi_{1} \\
\psi_{0} \\
\ea
\right)=P_{n}
\left(
\bea{c}
\psi_{1} \\
\psi_{0} \\
\ea
\right)\nonumber
\eeq

with $\psi_{n}$ the component of wavefunction
for energy E at site n

\beq
M_{n}=
\left(
\bea{cc}
0 &
-{\displaystyle {t_{n-1}}\over{\displaystyle t_{n}}}\\
{\displaystyle 1} & {\displaystyle 0}\\
\ea
\right)
\ \ \ \ \ \hbox{and} \ P(n)={\displaystyle \prod_{i=1}^{n}} M_{i}
\nonumber
\eeq


Connecting the system with external perfect metallic leads, one 
undertakes a transmission study that give access to the Landauer 
resistance  $\rho_{N}=h/e^{2}\ R/T$ 
where $T$ is the fraction of tunneling electrons transmitted from the 
system to the leads, and $R$ is the reflected one. It is directly related with the total transfer matrix 
elements, $\rho_{N}=\frac{1}{4}(P^{2}_{N}(1,1)+P^{2}_{N}(1,2)
+P^{2}_{N}(2,1)+P^{2}_{N}(2,2)-2)$ and in quasiperiodic Fibonacci 
chains, resistance can be written down in a closed form at energy E=0.

$$\bigl(\rho_{N}\bigr{)}_{\hbox{Fibonacci}}
=\frac{1}{4}(\gamma^{2s(N)}+\gamma^{-2s(N)})-\frac{1}{2} 
=\biggl(\frac{\gamma^{s(N)}-\gamma^{-s(N)}}{ 2}\biggr{)}^{2}$$

coefficients $s(N)$ following a quasiperiodic pattern (Fig.1). 
Phason disorder is introduced  via  -BB- units, resulting from of a transition 
between two isomorphic Fibonacci chains \cite{RocheFas}. One shows 
that for one phason, whatever its position, Landauer resistance for 
the N-th chain is exactly \cite{RocheFas}

$$\bigl(\rho_{N}\bigr{)}_{\hbox{one-phason}} 
=\Biggl(\frac{\gamma^{s(N)-1}-\gamma^{-s(N)+1}}{ 2}\Biggr{)}^{2}$$

\noindent
whereas in the highest density case corresponding to chains of 
typical form \\ (BB)-ABBABBAABABBABBAA-(BB) (where the external 
units are associated with constant hopping integrals) calculations 
lead to \cite{Roche-N}

$$\bigl(\rho_{N}\bigr{)}_{\hbox{multiple-phason}} 
=\Biggl(\frac{\gamma^{\tilde{s}(N)}-\gamma^{-\tilde{s}(N)}}{ 
2}\Biggr{)}^{2}$$

\noindent
with $\tilde{s}(N)$ a new complex function of N calculated 
iteratively which manifest self-similar fluctuations with respect 
to $s(N)$. From such studies one finds that phason disruptions of quasiperiodic order are not able to 
alter the localization mechanism which do remain basically the 
same in the limit $N\to\infty$. To unveil interesting aspects of 
quantum interferences one has to start from a continuous approach of 
the Schr\"{o}dinger equation in Kronig-Penney models in which the 
potential describing the 
interaction of the electron with the lattice is represented by a sum 
of Dirac distributions  with intensity $V_{n}$ localized at $x_{n}$. Following the works and discussions of Kollar and S\"uto\cite{Suto}, and
 M. Baake et al.\cite{Baake}, we focus our attention around so-called 
 conducting points ($k=k_s$), that correspond to the commutations of transfer 
 matrix, and which are worked out analytically 

$$\rho_{N}|_{_{_{k=k_s}}}=\biggl(\frac{\lambda}{2 k_s} \biggr)^2
\ \ \frac{\sin^2 N \varphi}{\sin^2 \varphi}$$

\noindent
with $\varphi$, depending on $k^{2}_{s}$ and $\lambda$, 
and is defined by $\cos \varphi=\cos k_s a+ \lambda/{2 k_s} \sin k_s a$. The 
analytical form of $\rho_{N}|_{_{_{k=k_s}}}$indicates that when $N\to\infty$, the 
Landauer resistance oscillates but remains bounded, so that the 
energies $k^{2}_{s}$ correspond to states that lead to best 
transmission, reminding that localized states will display exponential 
increase of resistance. For energies in the vicinity of $k_{s}$ 
$k^{2}=(k_{s}+\varepsilon)^{2}$  (taking ${\lambda}= 2 k_s (\cos 
\varphi_s-\cos k_s )/\sin k_s$ 
and ${\varphi}_{s}=(m \pi)/ N$ with m=1,...,N-1), one finds self similar patterns 
for values of the scattering potential around 
$\varphi\sim\frac{\pi}{2}$. This suggests that 
${\rho_{N}}\mid_{\hbox{one-phason}}(x_{P})$ reveal critical states 
which are robust against phason disorder as 
found in the TBM for E=0 case. The typical patterns represented on 
Fig.2 actually encloses oscillations of resistance which 
smaller oscillations are described by some coefficients $s(n)$ 
previously introduced.


Second, two symmetric zones corresponding to 
$ m <  N-4P, \ \hbox{and} \ m  >  4P$ are such that ${\rho_{N}}\mid_{\hbox{Fibonacci}}(x_{P},m,\varepsilon)>
{\rho_{N}}\mid_{\hbox{one-phason}}(x_{P},m,\varepsilon)$ in the 
strict sense ($x_{P}$ a degree of freedom related
 with the position of phason defects, P the total number of available positions
 for a given N). A simple pattern of ${\rho_{N}}\mid_{\hbox{one-phason}}(k,m,x_{P})\sim \alpha(m/P)\  
\sin\bigl( \frac{2m\pi}{P} x_{P}\bigr)$ is found with eigenfrequency of
 $m/P=5/472=0.0106$ as revealed by the power spectrum of m=5 curve is given 
 on Fig.3. Superimposed small oscillations are an unphysical effect due to a 
Fourier transform of a finite signal.


The different behaviors found in this study suggest that in some case 
local disruption of long range quasiperiodic order has improved the
 conductive ability of the chain in a systematic manner, which is 
 consistent with the abovementionned remarkable feature of 
 quasicrystalline materials. Analyzing the interference pattern of the Landauer resistance as a function of 
  phason defect suggest that extendedness (as a localization 
  properties of available states at such energies) has also been 
  jointly improved. This is shown by a bounded and simple oscillatory 
  pattern for the resistance, common to what is found for extended 
  eigenstates in a periodic systems.  Some affected states (by phasons) 
  may remain self-similar with same transmission ability as in the Fibonacci case whereas others will manifest an improved transmission ability.

\section{Quantum interference mechanisms in high-dimensionnal 
quasicrystals}

Some effort to investigate quantum interferences effects in small
quasiperiodic penrose approximants have been made \cite{Moulopoulos}. 
Here we propose how quantum interferences on 
both sides of a metal-insulator transition in real materials might be 
analyzed. Indeed, weak localization regime has been found in experiments for some 
quasicrystalline materials (AlCuFe,\ldots) whereas other systems such as 
AlPdRe-quasicrystals behave differently being very close 
to a metal-insulator transition\cite{Berger}. Two different focus may be considered for a general understanding of 
quantum interferences in quasicrystals. First, as there exist 
approximant phases (periodic) sharing the same behavior, weak 
localization correction beyond Drude approximation should apply for 
that systems as well as for corresponding quasicrystals. The 
only recurrent approximation when solving the cooperon diffusion equation 
is to assume that scatters are uncorrelated i.e
$\<{\cal U}(r){\cal U}(r')\>_{disorder}=cu^{2}\delta(r-r')$
(${\cal U}(r)= \sum_{i=1}^{N}\  {\cal U}(r-R_{i})$, c the impurity 
concentration, u typical strength). The calculation of 
the quantum correction of the conductivity in this regime is
 enclosed in phase factor interferences of the two-particle 
 Green's function  $\<GG^{*}\>$ . By performing configuration averaging beyond the mean free path legnth
scale, then $\< GG^{*} \>_{disorder}$ reduces phase interference to 
the ($k'=-k$)-Cooperon pole, as 
a consequence of time reversal symmetry, the possibility to have a 
coherent distribution of scatters is usually neglected.

However, assuming that the distribution of scatters is
constrained to, let's say for sake of illustration, a mirror-plane symmetry, i.e 
$\forall \alpha \in  \{ R_{\alpha}, \alpha=1,N\}$ there is a site
 $R_{\beta}$ such that $R_{\beta}=-R_{\alpha}$, then without performing any 
 diagrammatic expansion, we just notice that weak localization related with average of the potential 
 scattering $\< {\cal U}(r){\cal U}(r'){\>}_{disorder}$ corresponding to
 phase factors {\small $\sum_{\alpha\beta}\< {e}^{-i(k\ R_{\alpha}+p\ 
R_{\beta})}{\>}_{disorder} =\frac{1}{\Omega^{2}}\ \int d^{3}R_{\alpha}d^{3}R_{\beta} {e}^{-ikR_{\alpha}}
\ {e}^{-ipR_{\beta}}$} will display new terms associated to above-mentionned symmetry, 
$\< {e}^{-i(k\ R_{\alpha}+p\ R_{\beta})} {\>}_{disorder}=
 \< {e}^{i(p-k) R_{\alpha}} {\>}_{disorder}=\delta_{p-k,0}$ that will 
 increase the contribution of phase interferences.  
 Say in another way, if a double symmetrical loop crosses in the mirror plane and in a region with extension less or equal to $\lambda_{F}$, then four equivalent pathes 
will interfere at the returning point instead of the usual two of the weak localization scheme, resulting in a total interference
  amplitude will be 4 times stronger ($\mid {\cal A}_{I}+{\cal A}_{II}+{\cal A}_{III}+{\cal A}_{IV}\mid 
  ^{2}=16\mid {\cal A}\mid ^{2} $). Similar
   ideas have been already introduced in context
    of mesoscopic physics \cite{Baranger} but we propose here that 
    they may serve as a path to follow a metal-insulator
     in quasicrystals in which even disorder may keep
      some strong correlated properties\cite{Roche-un}.

The second focus is suggested by the close proximity of a metal-insulator 
transition and gives a very dissimilar weight to quantum interferences. Any critical point of an electronic localisation-delocalisation transition
can essentially be described by its anomalous diffusion which means that two-particle Green function reads
 $\< |G^{+}(r,r';E)|^{2}\>\sim|r-r'|^{-\eta +2-D}$ with $\eta$ a critical 
 exponent (and the propagator representes the transition 
 probability in real space of an electron of E energy from site
$|r\>$ to $|r'\>$), and which directly affect the conductivity of the
system since 

$$\sigma_{DC}=\frac{2e^{2}}{h}\lim_{\varepsilon\to 0} 
4\varepsilon\int d^{D}r r^{2}\< |G^{+}(r,r';E)|^{2}\>_{disorder}$$

Power law behavior have been found analytically in 
quasiperiodic systems\cite{Jona,Bell}, numerically 
for quantum Hall systems \cite{QHE} and 3D quasiperiodic systems
\cite{RocheJMP,RochePRL1}. No characteristic length scale can be defined (no more 
validity of mean free path approximation) and interference effects are 
intrinsically defining the dominant transport mechanism. This has 
also been recently discussed on realistic models of 3D-quasicrystals by 
scaling analysis of the Kubo-formula \cite{RocheFS}.

In both cases the presence of phasons as demonstrated in 1D systems 
may weeken the interference effects while destroying quasiperiodic 
long range order. Working this out analytically in realistic models remains a great 
challenge.

\stars
S.R. is indebted to the European Commission for financial support 
(Contract {\small ERBIC17CT980059}), and to Prof. T. Fujiwara from Department of Applied Physics of
Tokyo University for his hospitality. Kostas Moulopoulos and Udo Beierlein 
are acknowledged for encouragements.

\vfill\eject


Figure captions:

\vspace{10pt}

\noindent
Fig.1 : Multifractal distribution of $|s(N)|$ for the Fibonacci chain of 800 sites.

\noindent
Fig.2 : The power spectrum of a self-similar pattern shown in the inset. 
The highest frequency is given by $\nu=0.5$ which is related to the change of $\rho_{N}(x_{P}\to 
x_{P}+1)$. On respective figures five unambiguous frequencies have been located and 
named $\nu_{n=1,5}$

\noindent
Fig.3 : Power spectrum of the pattern given in the inset for m=5, chain of 
2000 sites.



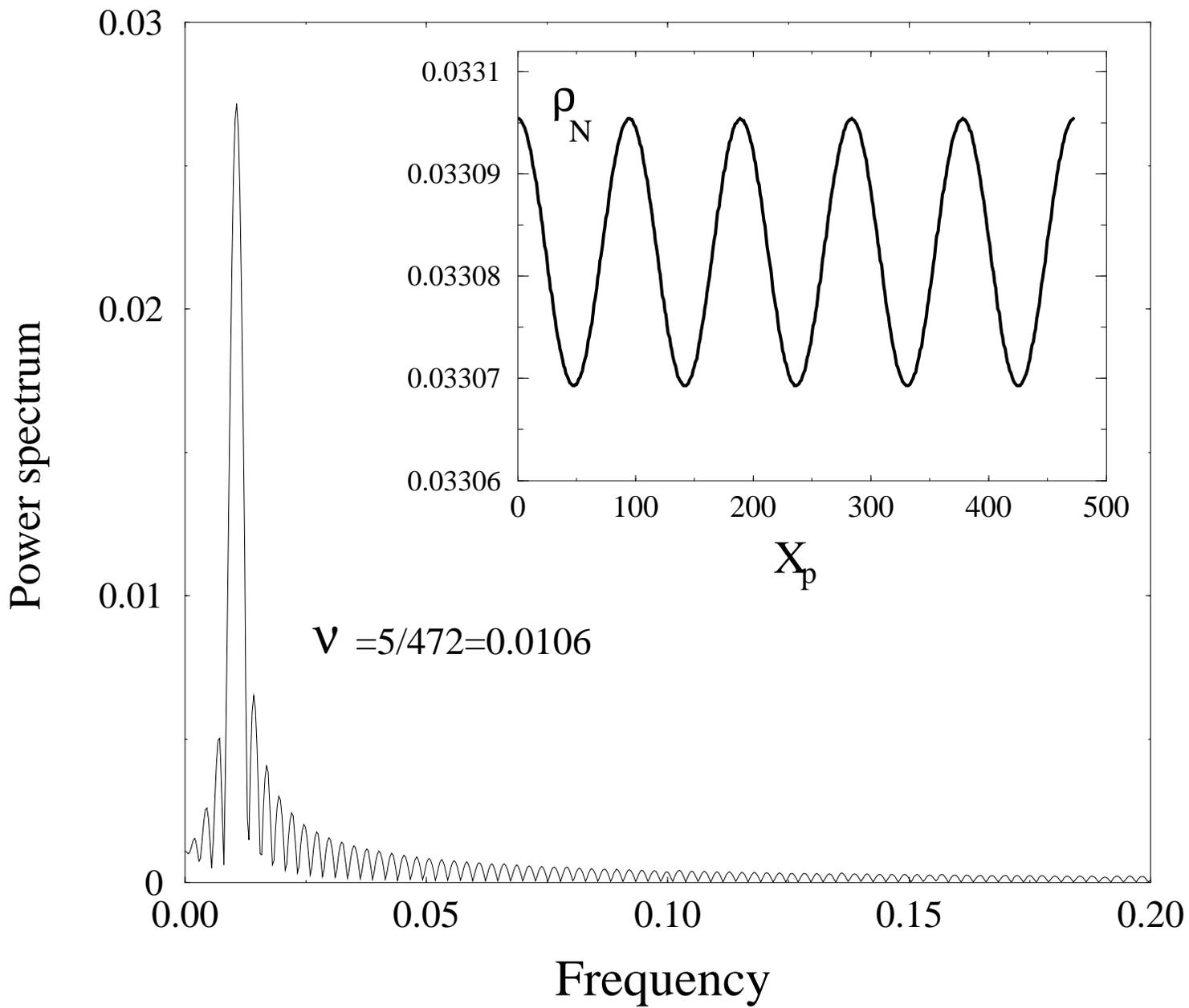
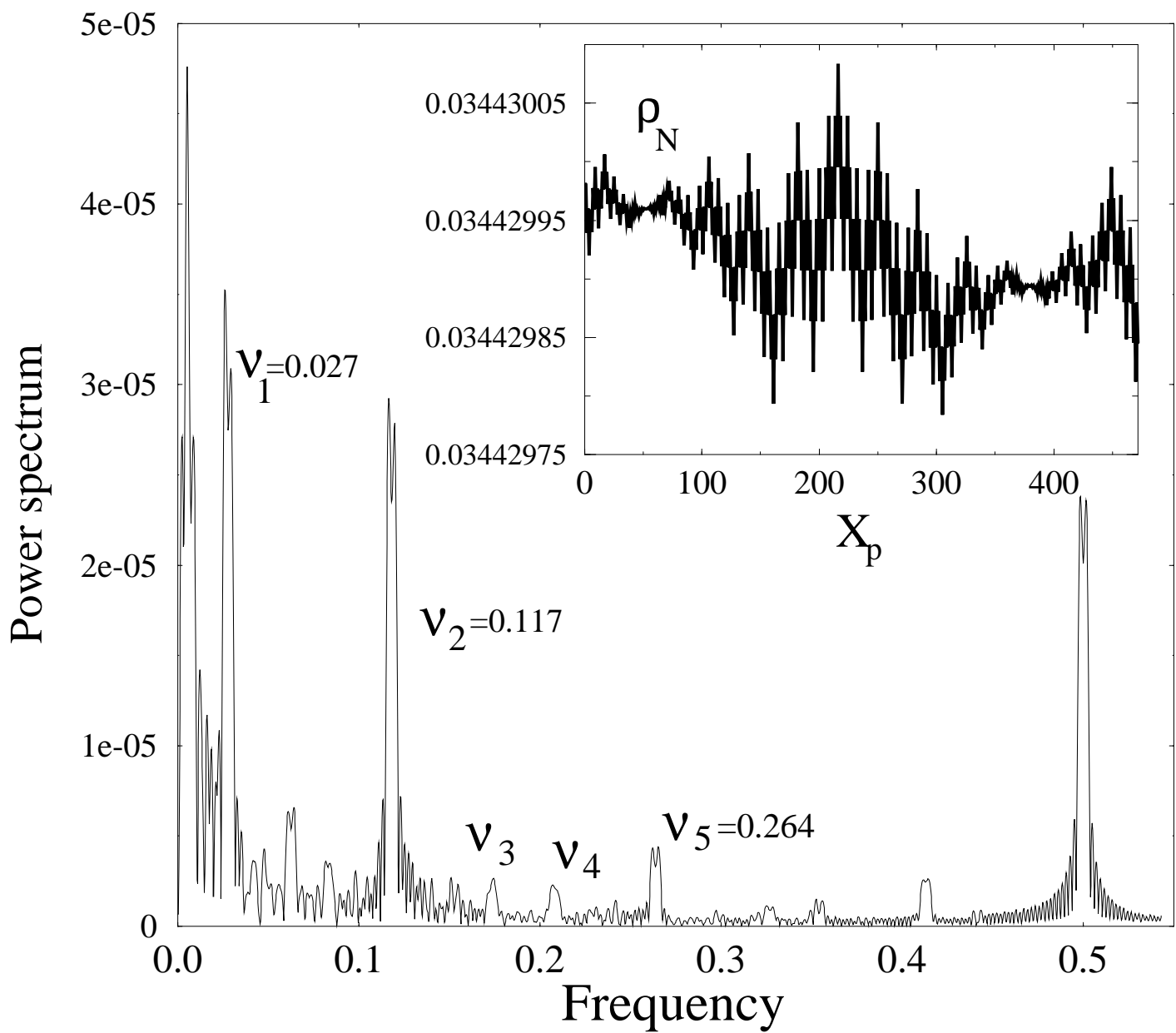
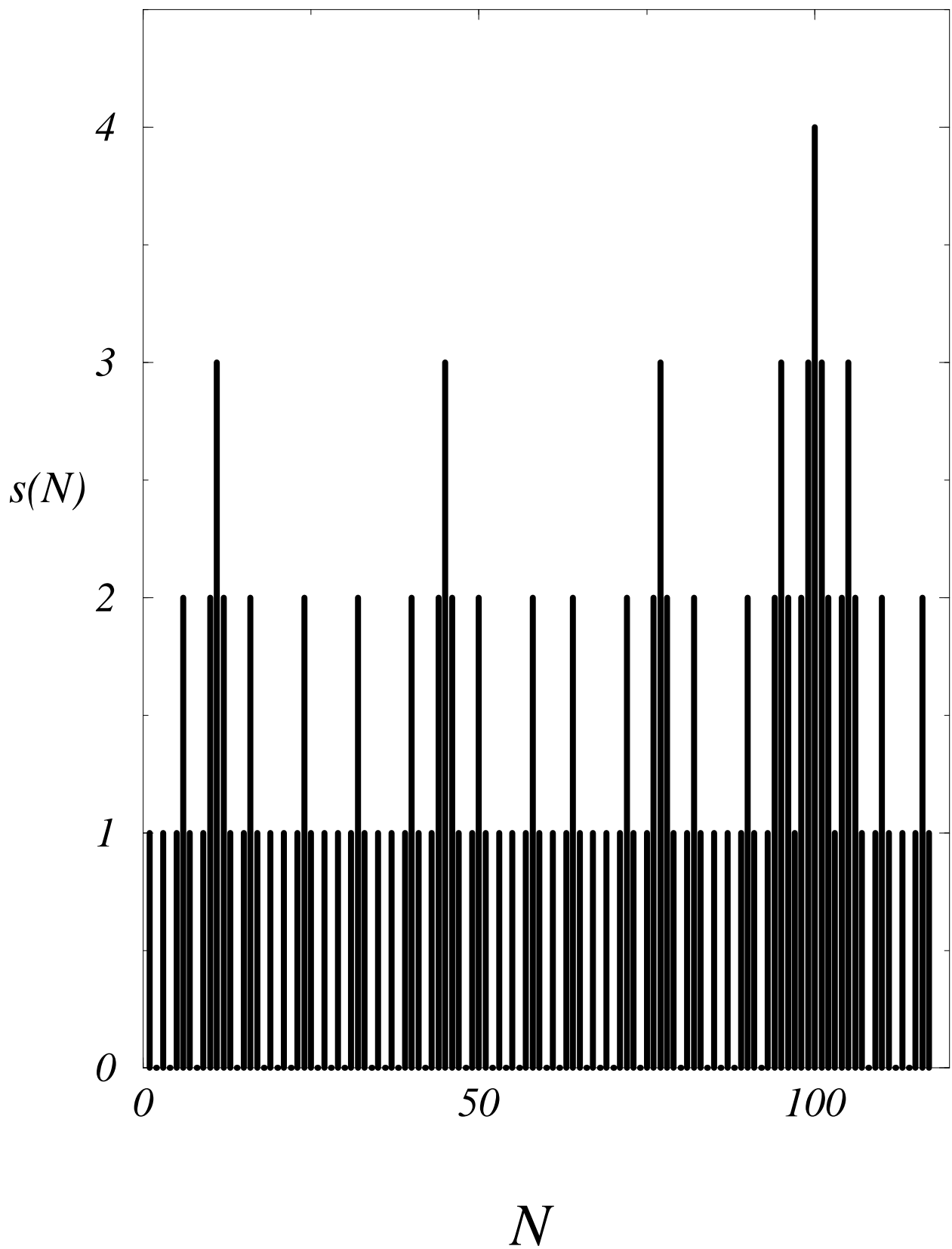
\end{document}